\documentclass[pdflatex,sn-chicago]{sn-jnl}

\usepackage{graphicx}%
\usepackage{multirow}%
\usepackage{amsmath,amssymb,amsfonts}%
\usepackage{amsthm}%
\usepackage{mathrsfs}%
\usepackage[title]{appendix}%
\usepackage{xcolor}%
\usepackage{textcomp}%
\usepackage{manyfoot}%
\usepackage{booktabs}%
\usepackage{natbib}
\usepackage{algorithm}%
\usepackage{algorithmicx}%
\usepackage{algpseudocode}%
\usepackage{listings}%

\theoremstyle{thmstyleone}%
\theoremstyle{thmstyletwo}%

\theoremstyle{thmstylethree}%

\begin{document}

\title[Article Title]{An Empirical Study on how Computing Students Interact with Large Language Models for Learning Purposes}


\author[1]{\fnm{Opetunde} \sur{Ibitoye}}\email{ibitoyoo@mail.uc.edu}

\author[1]{\fnm{Saheed} \sur{Popoola}}\email{saheed.popoola@uc.edu}

\affil*[1]{\orgdiv{School of Information Technology}, \orgname{University of Cincinnati}, \orgaddress{\city{Cincinnati}, \postcode{45221}, \state{Ohio}, \country{USA}}}


\abstract{Large Language Models (LLM) have become popular tools in education. Therefore, understanding how students interact with LLMs for learning purposes is an important aspect of the responsible design and development of LLM-based learning tools. However, existing research has focused on evaluating model performance, specific use cases, and students’ attitudes toward LLM in education. There has been limited empirical work exploring students' interaction strategies with LLMs in computing education. This study draws on Self-Regulated Learning (SRL) theory and conducts a survey of 225 undergraduate and graduate computing students to investigate how students use LLM tools during learning, the perceived learning impact, and the strategies they rely upon when using these tools. Students used LLMs across nine functional categories, most often to explain concepts, learn new material, find information, debug, and generate ideas. They also adopted some strategies such as supplying detailed context and iterative prompting. The most common difficulties encountered with LLM tools were incorrect information, generic responses, and difficulty trusting the model’s reasoning. We synthesized these patterns through an SRL lens and offer a Probe–Monitor–Shape heuristic to map students’ LLM interaction strategies onto the phases of self-regulated learning.}




\maketitle
\section{Introduction}


The characterization of how students interact with Large Language Models (LLM) is essential for designing responsible LLM-based learning tools that align with pedagogical intentions and anticipate how LLMs reshape students’ cognitive and metacognitive processes \citep{razafinirina2024pedagogical, narreddy2025harnessing}.
Researchers have examined how computing students interact with LLMs by conducting surveys and analyzing usage or conversation logs, reflections, and self reports \citep{EstelleSmith2024, Ma2024, shin2024understanding, Frankford2024, alpizar2025student}. Others have evaluated model performance and conducted short and long term studies on human subjects to categorize student-LLM interactions \citep{Lyu2024, Neyem2024, Bassner2024, Li2024}.
However, many of these studies only examined the performance of a specific tool within specific course contexts or programming tasks.  
These studies describe observable actions without capturing the underlying intentions or strategies that characterize these interactions. This can lead to fragmented and context-bound insights that cannot be generalized across different learning environments. 
This fragmentation can limit our ability to detect emerging challenges, interpret learning impacts, and design interventions that respond to the realities of student–LLM interaction.

To address this gap, this study examined the purposes for which computing students use LLMs, the needs that motivate their use, the strategies they employ during their interactions with LLMs, and their perceived learning impact. We then developed a taxonomy of LLM use and strategies in computing education based on students’ self reported practices. We adopted the Self-Regulated Learning (SRL) theory \citep{zimmerman2002becoming} as a conceptual lens to interpret these interactions in a principled way. SRL characterizes learning as a cyclical process in which learners set goals and plan (forethought), monitor and control their strategies while working (performance), and evaluate and adapt afterward (self-reflection). The view of student–LLM interaction as self-regulated learning process provides a student-centered basis for a taxonomy of the purposes, strategies, and appraisals that structure LLM use in computing education.
The research questions guiding this study are as follows:

\begin{itemize}
\item {\texttt{RQ1.}} How do students use LLM tools to support their academic learning?
\item {\texttt{RQ2.}} What strategies do students employ when interacting with LLMs tools for learning?
\item {\texttt{RQ3.}} What perceived benefits and challenges do students report regarding the use of LLM tools for learning?
\end{itemize}

This study contributes to the body of literature on the use of LLMs in computing education in two ways. First, it offers an empirically grounded taxonomy of student–LLM interactions that characterizes how students integrate LLMs into their learning processes. Second, it provides a structured account of the benefits and challenges that students perceive when using LLMs. Third, it introduces Probe–Monitor–Shape, a heuristic that organizes students-LLM interaction strategies and maps them onto the phases of self-regulated learning. The heuristic is offered as a preliminary organizing concept and a starting point for instructional support and future empirical work.
Together, these contributions provides a foundation for designing learning environments, instructional support, and institutional policies that respond to the evolving role of LLMs in the learning process.


\section{Background and Related Work}
\subsection{LLMs in Computing Education}
LLM-based tools have evolved to become part of students’ learning environments and are used by students at different educational levels for various tasks across computing courses \citep{EstelleSmith2024, raihan2025large}. 
Research on the implications of commercial LLM tools such as ChatGPT, Gemini, Copilot, and Claude shows their potential to support learning \citep{Bttcher2025}. Custom LLM tools have demonstrated efficacy in supporting learning by functioning as supports to assist in teaching and learning processes \citep{Lyu2024, Mueller2025, Bassner2024}. Similarly, other studies have introduced LLM-based systems to monitor and assess student performance \citep{Tang2024, Li2024}. These systems enable instructors to oversee student learning behavior in real time.
The broad applications of LLMs demonstrate their potential to enhance traditional computing education, and foster personalized learning and engagement \citep{Abolnejadian2024}. However, challenges such as overreliance on LLM, inappropriate help-seeking, misinformation, and LLM literacy levels may negatively impact learning and knowledge retention in computing classrooms \citep{Kazemitabaar2023, Hou2024}.
The body of literature examined above show that LLM tools have become routine components of the teaching and learning processes in computing education. Therefore, it is important to understand how students incorporate LLM tools into their learning processes. This understanding will help align LLM tools with learning objectives, pedagogical goals, ethical constraints, and proper support calibration. The understanding can also help identify whether systematic interventions are needed. This will facilitate the responsible use of tools by students and maximize the learning impact of LLMs in computing education for students.


\subsection{Student–LLM Interaction in Computing Education}
Understanding how students incorporate LLMs into their learning processes has become increasingly important as LLM tools have moved from peripheral aids to routine components of learning in computing education. Research shows that active engagement between students and LLM tools in computing education is largely influenced by the generative capabilities of LLMs. Students use LLMs tools to understand, learn, and perform various tasks such as content generation, text summarization, creative brainstorming, and coding \citep{shin2024understanding}.
Furthermore, research suggests that LLM tools significantly influence how students learn computing topics through an interaction driven process \citep{alqarni2026explicit}. The educational outcomes in this process are primarily mediated by specific textual prompting patterns and the student's ability to iteratively refine these interactions to scaffold their own learning \citep{gong2025exploring, shi2025large}. Across these interaction contexts, students interpret model outputs, evaluate their relevance, adjust their strategies, and compensate for system limitations as part of their learning processes.


Most of the existing research examines students in a specific context of use or academic level. This pattern of evidence can obscure more granular interactions, limit broader applicability, and limit the detection of systematic student-LLM interaction patterns for learning. This scenario creates a critical gap between the tool’s technical capabilities and the social realities of its use for learning. 
This study extends the computing education literature by providing an empirical account of students’ real world interactions with LLMs. The study addresses a gap in how the field conceptualizes and measures the processes that underlie LLM‑supported learning.

\subsection{Self-Regulated Learning}
Self-Regulated Learning (SRL) offers an established framework for interpreting how learners direct their own study. In Zimmerman’s cyclical model, regulation unfolds across three phases: a forethought phase of task analysis, goal-setting, and strategic planning; a performance phase of self-control and metacognitive self-observation while the task is carried out; and a self-reflection phase of self-judgment and adaptation that feeds back into the next cycle \citep{zimmerman2002becoming}. Effective learners move through these phases repeatedly, adjusting strategies as they gauge their progress against a goal.
Within this framework, help-seeking is itself a self-regulatory strategy rather than a sign of dependence. Research distinguishes instrumental help-seeking, in which learners request the minimum support needed to continue reasoning on their own, from executive help-seeking, in which learners offload the task and its cognitive work onto the source of help \citep{karabenick2013help}. The distinction is consequential as instrumental help-seeking supports learning, but executive help-seeking can undermine it. Prior work in computing education has examined how generative AI reshapes students’ help-seeking preferences \citep{Hou2024}, but has rarely connected these choices to the phase structure of self-regulation. We adopt SRL and the instrumental-executive distinction as the lens for this study. We use the framework to interpret the strategies students report, to explain uneven verification behavior as an imbalance across regulatory phases, and to ground the Probe–Monitor–Shape heuristic introduced in Section \ref{discussion}.




\section{Methodology}
\subsection{Research Design}
This study used a mixed methods survey design to characterize how computing students use LLM tools for learning, the strategies they employ, and the benefits and challenges they perceive. We draw on Self-Regulated Learning (SRL) theory \citep{zimmerman2002becoming} as a conceptual lens to ground the design and interpretation of the study. SRL frames learning as a cyclical process in which learners set goals, monitor their progress, and adapt their strategies. This lens informed the construction of the survey instrument, the development of the qualitative codes, and the interpretation of the resulting categories. We treat the data as a descriptive snapshot of self-reported practice rather than as evidence of causal effects.


\subsection{Survey Instrument}
A questionnaire of open and close-ended questions was developed to capture students’ use and perceptions of LLM tools in computing education. The instrument was written in English and organized into three sections aligned with the study objectives. The instrument include three demographic items, six items on learning and use, and fourteen items on perceived benefits, challenges, and strategies. Item formats included single and multiple select questions, five-point Likert items, and free text responses. Each item was reviewed for relevance and clarity by the authors, and pilot-tested with 10 students prior to deployment. 

\subsection{Participants}
Participants were students enrolled in undergraduate and graduate computing-related programs such as computer science, information technology, computer engineering, information science, and cybersecurity, across multiple academic levels and geographic regions. 32.0\% of the respondents identified as female, 50.2\% as male, 0.5\% as nonbinary, 2.2\% preferred not to say, and 15.1\% left the item blank. By field, 33.96\% majored in Information Technology, 30.19\% in Computer Science, 3.77\% in Cybersecurity, 2.36\% in Computer Education, 1.89\% in Computer Engineering, and 1.89\% in Information Science; 8.49\% came from fields outside computing, and 17.45\% did not report a field. By level, 48.9\% were undergraduates, 21.3\% were master’s students, 14.2\% were PhD students, and 15.6\% did not report an academic level.

\subsection{Data Collection}

The survey was administered anonymously and voluntarily via Qualtrics from February 17, 2026 to April 26, 2026. Participants were recruited publicizing on social media, at computing conferences, and sending mails to heads of computing units at university. We prioritized academic units with large number of computing students, these universities were found on the ranking site. After data cleaning, in which duplicate, empty, and substantially incomplete responses were removed, a final analytic sample of 225 responses was retained. Because most items were optional, the number of valid responses varies by item; we therefore report the valid N alongside each analysis rather than assuming a uniform denominator. In particular, 166 participants completed the Likert items, 102 answered the open-ended strategy item, 103 answered the open-ended challenge item, and 107 answered the open-ended benefit item.




\subsection{Ethical Considerations}
This study was reviewed and approved by the Institutional Review Board (IRB) of the authors’ institution. Participation was voluntary and fully anonymous, and no personally identifying information was collected. Before beginning the survey, participants reviewed an information sheet describing the purpose of the study, the voluntary nature of participation, and how the data would be stored and used, and they provided informed consent. Participants could skip any item or withdraw at any point without penalty. 

\subsection{Data Analysis}

The data were analyzed using a mixed methods approach aligned with the structure of the instrument, and the research questions were mapped to their corresponding survey items Table \ref{tab:rq_survey_mapping}. Closed-ended and Likert items were analyzed with descriptive statistics. Likert responses were first converted to a 1–5 numeric scale. Quantitative analyses were conducted in Microsoft Excel and Python. 

Open-ended responses were analyzed using conventional (inductive) content analysis \citep{hsieh2005three, stemler2001introduction}. Coding proceeded in three passes. In the first pass, two researchers independently read the full set of responses for each open-ended item and generated initial open codes directly from the data. In the second pass, the researchers met to compare codes, reconcile overlaps, and consolidate them into a shared codebook, with each category given an explicit definition and anchor examples. In the third pass, both researchers applied the finalized codebook to the complete set of responses. Coding was not mutually exclusive: a single response could express more than one purpose, strategy, or challenge and was assigned all applicable codes, so reported counts reflect the number of mentions rather than the number of respondents. Coding disagreements were resolved through discussion to consensus. Inter-coder agreement on an independently double-coded subset had a Krippendorff's alpha score \citep{krippendorff2011agreement} of 0.85. 
Finally, the categories were organized under the three research questions. 

\begin{table*}
\small
  \caption{Mapping of Research Questions to Survey Questions}
  \label{tab:rq_survey_mapping}
    \begin{tabular}{p{6cm}p{9cm}}
    \toprule
    Research Question & Survey Questions \\
    \midrule
    \textbf{RQ1:} How do students use LLM tools to support their academic learning? & 
    Q1: What do you use LLM tools for? \newline 
    Q2: What academic task do you use LLM tools for? \newline
    Q3: Please describe a recent situation where you used an LLM tool for learning. \\
    \midrule
    \textbf{RQ2:} What strategies do students employ when interacting with LLM tools for learning? & 
    Q4: Which strategies do you typically use when interacting with LLM tools?  \newline 
    Q5: Please describe one strategy you have found most effective for getting useful results from an LLM tool. \\
    \midrule
    \textbf{RQ3:} What perceived benefits and challenges do students report regarding the use of LLM tools for learning? & 
    Q12: Using LLM tools has helped me understand computing concepts more clearly? \newline 
    Q13: I feel that LLM tools save me time on tasks related to my coursework. \newline 
    Q14: LLM tools make me feel more confident when learning independently? \newline 
    Q16: I worry about becoming too dependent on LLM tools? \newline 
    Q17: Technical or access barriers (e.g., cost, login restrictions, slow responses) limit my use of LLM tools. \newline
    Q18: LLMs risk weakening my long-term understanding of concepts? \newline 
    Q19: I have received incorrect or misleading answers from LLMs? \newline 
    Q20: I sometimes copy or rely on LLM output without fully understanding it? \newline 
    Q23: Describe a situation where using an LLM tool created difficulties or confusion in your learning? \newline
    Q24: In your experience, what has been the most valuable benefit of using LLM tool for your learning \\
    \bottomrule
  \end{tabular}
\end{table*}

\section{Results}


\subsection{How do students use LLM tools to support their academic learning?}
The results of the analysis showed that students used LLMs to support their learning across nine functional categories. These categories are explaining or clarifying concepts (n = 31), learning new concepts (n =8), brainstorming ideas for assignments or projects ( n = 10), research support (n = 14), writing support (n = 9), exam preparation and practice (n = 10), debugging or troubleshooting errors (n = 15), project development or technical implementation (n = 6), and checking logic or verifying solutions (n = 2). 

\subsection{What strategies do students employ when interacting with LLM tools for learning?}

Of the 102 students who answered the open-ended strategy item, 90 described a codable strategy, producing 111 strategy mentions across seven categories. The most frequently reported strategy was providing detailed context to constrain the model’s response (n = 46), followed by iterative prompting (n = 22), constrained prompting (n = 13), and verifying outputs against answer keys, external sources, or the student’s own reasoning (n = 16). Less frequent but recurring strategies included prompting the model to ask for the information it needed (n = 7), combining model output with other sources (n = 4), and setting an explicit goal before prompting (n = 3). Table \ref{tab:interaction_strategies} provides a summary of the categories of interaction strategies reported. In the table, counts reflect mentions and are not mutually exclusive. 

\begin{table*} 
\caption{Categories of interaction strategies reported.} 
\label{tab:interaction_strategies} 
\begin{tabular}{p{0.25\textwidth} p{0.30\textwidth} p{0.30\textwidth} p{0.08\textwidth} c} 
\toprule
\textbf{Category} & \textbf{Description} & \textbf{Example Quote} & \textbf{Count}\\ 
\midrule 

Provide detailed context & Supply background information, assignment instructions, drafts, outlines, or source materials. & ``Giving a lot of context-the draft of my paper, an outline, the sources I want to use. & 46 \\ 
\hline
 Constrained prompting & Giving LLMs specific format or rules to think by and produce response based off. & ``Explain this code step-by-step for a beginner coder.'' & 13 \\ 
\hline
 Goal setting before prompting & Defines the desired outcome in advance in order to detect drift.  & ``Understanding my required outcome before the LLM produces output for me so I can easily detect when it is veering off.'' & 3 \\ 
\hline
Iterative prompting &   Reshaping questions, ask clarifying questions, or try different phrasing. & ``Continuously reshape questions until I get the response I want.'' & 22 \\ 
\hline
 Asking LLM needs & Asking the tool what information it re-quires to respond effectively. & ``Ask the LLM about what it needs to assist effectively'' & 7 \\

 \hline
 Combining sources & Integrating LLM outputs with other LLMs, personal knowledge or external materials. & ``Combining different sources and my own baseline knowledge together.'' & 4 \\ 
\hline
Verification & Using the LLM to confirm correctness or list relevant concepts. & ``I like to use it to check answers when no answer key is provided.'' & 16 \\ 
 
\bottomrule 
\end{tabular} 
\end{table*}




\subsection{What perceived benefits and challenges do students report regarding the use of LLM tools for learning?}
\subsubsection{Perceived Benefits of LLM Use}
\begin{figure}
    \centering
    \includegraphics[width=1\linewidth]{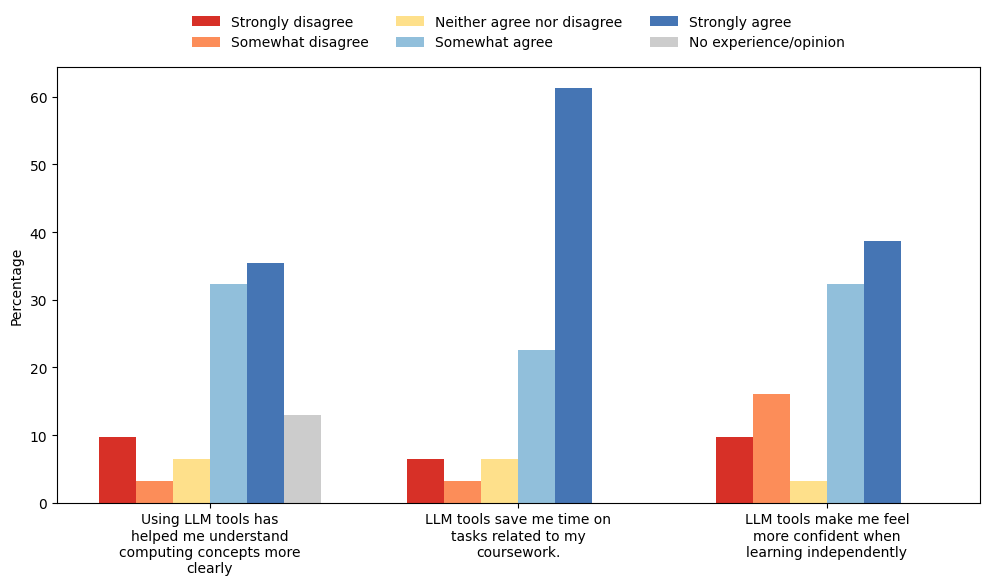}
    \caption{Percentage distribution of benefits of LLM use for Learning}
    \label{fig:bene}
\end{figure}

Most students reported that LLM tools help them to understand computing concepts more clearly, while some others reported using LLM tools because it saved time on tasks related to their coursework. A few students used LLM tools to make them feel more confident when learning independently. Responses to the three items assessing the benefits of LLM tools for learning are presented in Figure \ref{fig:bene}.

Open-ended responses provided more details on the benefits. 107 students answered the open-ended benefit item, and we extracted 91 benefit mentions across four major categories. By a wide margin, the most common benefit was concept explanation and simplification, with students describing LLMs as tools that broke down complex or unfamiliar material into simpler, more personalized explanations than lectures or textbooks provide (n = 45). Students also frequently valued LLMs for research, summarization, and information sourcing (n = 23). Other cited benefits are help coding, identifying or debugging errors (n = 11) and around-the-clock availability (n = 12). A small number of responses (n = 16) contained no codable benefit, either explicit non-use ("I don't use LLMs for my learning"), single-word or ambiguous answers, or explicit statements that the tool provided no benefit.

\subsubsection{Challenges Students Encounter When Using LLM Tools}
\begin{figure}
    \centering
    \includegraphics[width=1\linewidth]{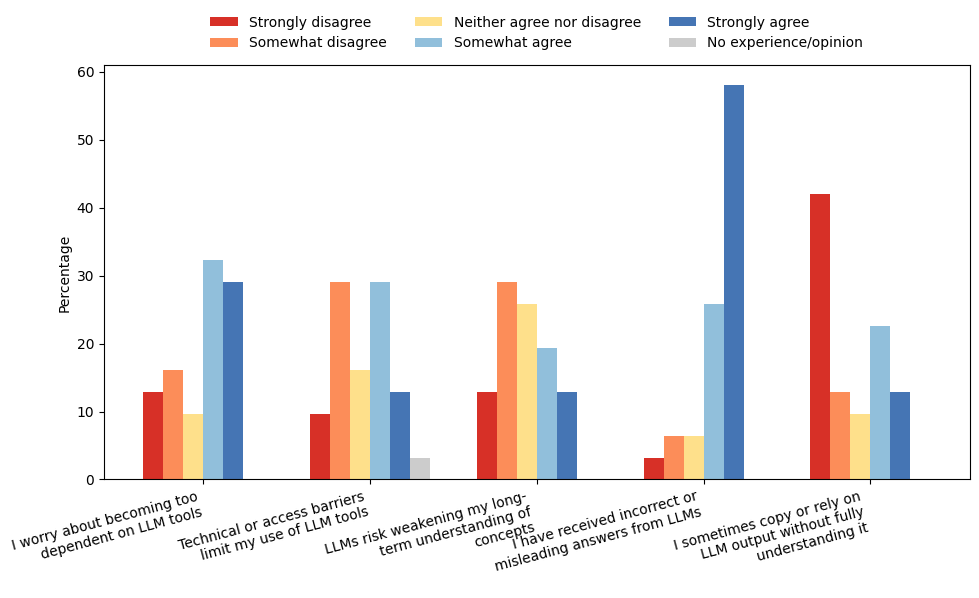}
    \caption{Challenges of LLM tools for learning.}
    \label{fig:cha}
\end{figure}

Students reported several difficulties when they use LLM tools for learning. The most common issue was receiving incorrect or confusing answers as noted by 83.87\% of the respondents. Concerns about becoming too dependent on LLMs were also frequent, with 61.29\% of the respondents agreeing or strongly agreeing.
Views on long term understanding were split: 32.25\% felt LLMs could weaken their conceptual grasp, while 41.93\% disagreed. Technical or access problems were reported by 38.71\% of students, with similar disagreement levels. Experience relying on LLM output without fully understanding it was also divided: 41.94\% strongly disagreed, and 35.48\% agreed or strongly agreed. Responses to the three items assessing the challenges of LLM tools for learning are presented in Figure \ref{fig:cha}.

Qualitative responses added detail to these challenges. Of the 103 students who answered the open-ended challenge item, 102 described a codable difficulty, producing 106 challenge mentions across seven categories. By a wide margin, the most common challenge was receiving incorrect, fabricated, or contradictory information (n = 46). Students also frequently reported generic or off-target responses that drifted from the prompt or ignored the course context (n = 31), and difficulty trusting or auditing the model’s reasoning such as when the LLM explanations are inconsistent or unclear (n = 16). Less frequent challenges included output that was too complex for the task or instructional level (n = 5), domain restrictions such as when LLM refuses to help due to safety restrictions (n = 3), technical or performance limitations (n = 3), and self-reported over-reliance (n = 2). This qualitative pattern is consistent with the Likert data, in which receiving incorrect or misleading answers was the most widely endorsed concern.

\begin{table*}[t]
\centering
\caption{Mapping of Probe-Monitor-Shape strategies to the challenges they
address and the benefits they help realize.}
\label{tab:mapping}
\small
\renewcommand{\arraystretch}{1.25}
\setlength{\tabcolsep}{3pt}

\begin{tabular}{@{}
p{0.20\textwidth}
p{0.23\textwidth}
p{0.27\textwidth}
p{0.27\textwidth}
@{}}
\toprule
\textbf{P-M-S element} &
\textbf{Strategies} &
\textbf{Challenges addressed} &
\textbf{Perceived benefit realized} \\
\midrule

\textbf{Probe}\newline
\emph{test \& explore what the model can do} &
Iterative prompting ; asking LLM needs; goal setting before prompting &
Generic responses; over-complexity. &
Greater confidence in independent learning; coding \& debugging; saves time. \\

\addlinespace

\textbf{Monitor}\newline
\emph{verify \& cross-check output} &
Verification; combining sources. &
Incorrect or confusing answers; difficulty to audit the model reasoning; over-reliance on the tools &
Research \& information sourcing; clearer conceptual understanding; greater confidence in independent learning \\

\addlinespace

\textbf{Shape}\newline
\emph{constrain \& steer the output} &
Provide detailed context; constrained prompting. &
Generic responses, difficulty to trust model reasoning, over-complexity. &
Clearer conceptual understanding; greater confidence in independent learning; coding \& debugging. \\

\bottomrule
\end{tabular}
\end{table*}

\section{Limitation and Threats to Validity}
The findings presented in this paper represent a snapshot of computing students’ experiences with LLM tools during the data collection period. The sample was limited to students enrolled in computing related programs and did not include students in other fields or educational sectors. The study relied on self reported survey data and primarily descriptive analyses, which may not fully capture students’ actual behaviors. In addition, LLM tools and institutional practices are rapidly evolving. Therefore, the results should be interpreted in the context of a specific timeframe.

\section{Discussion and Conclusion}\label{discussion}
LLM tools are reshaping the learning process in computing education by shifting students’ work from retrieving information to managing the behavior of an unpredictable system. Students engaged LLMs for explanations, debugging, idea generation, and validation of their understanding of a concept. However, student interactions were consistently shaped by the need to evaluate and correct responses that were incomplete, imprecise, or misaligned with the course expectations. This shift reframes help-seeking as a process that requires students to regulate tools rather than simply consume their output. These findings extend prior work showing that students often rely on inaccurate model responses and may not consistently verify them \citep{balse2023evaluating, elsayed2024impact}.

Across the strategies in Section 4.2, students’ self-reported practices clustered into three recurring regulatory moves that we summarize as Probe–Monitor–Shape. Interpreted through Zimmerman’s cyclical model of self-regulated learning \citep{zimmerman2002becoming}, the three moves correspond to distinct phases of regulation. Probing operationalizes the forethought phase where students set goals and plan how to approach an unfamiliar tool through goal-directed and iterative prompting and by asking the model what it needs (n = 22, 7, and 3). Monitoring corresponds to metacognitive self-observation, as students verify and cross-check output against a standard (verification and combining sources, n = 16 and 4). Shaping reflects performance-phase self-control, as students adapt the tool through detailed context and constrained prompting (n = 46 and 13). The loop we observe is therefore the cyclical structure of self-regulation enacted through an external system, which gives the coherence of the loop a theoretical basis.

We also noted that students engaged the control-oriented moves of shaping far more than the self-observation of monitoring (context-supplying, n = 46, versus verification, n = 16). This imbalance in which students steer the tool effectively but under-verify its output is precisely the dynamic that underlies the over-reliance and long-term understanding concerns students reported. We therefore present Probe–Monitor–Shape as a heuristic for organizing these behaviors and locating them within the SRL cycle, and not as a validated theoretical model.

\begin{figure}
    \centering
    \includegraphics[width=1\linewidth]{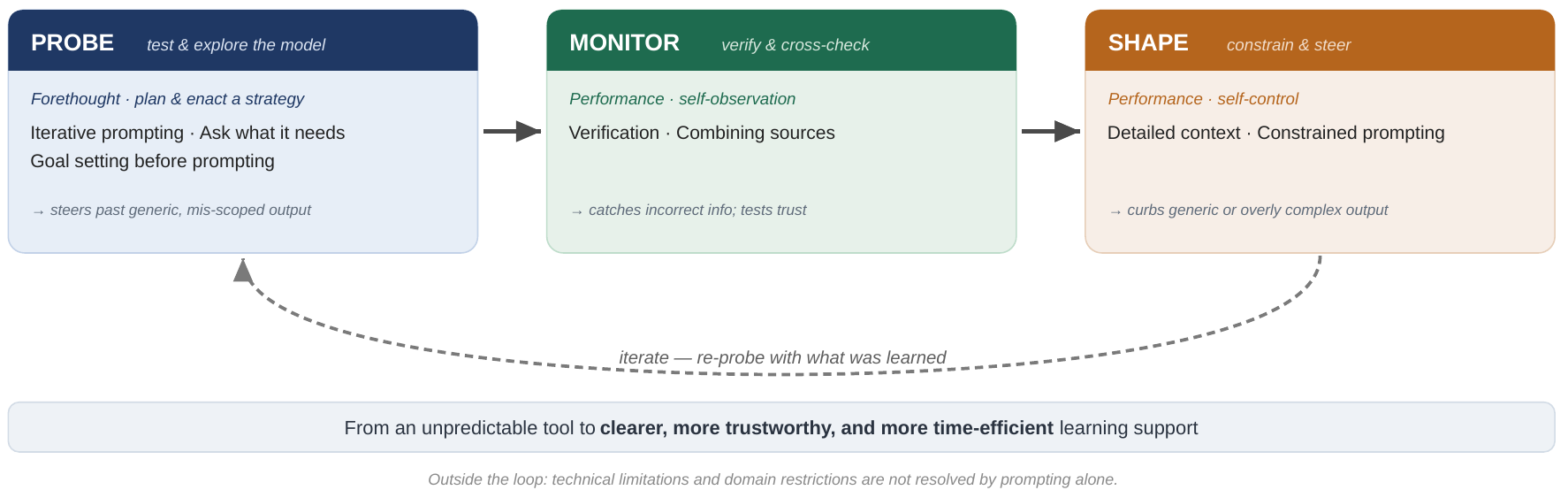}
    \caption{The Probe-Monitor-Shape Heuristic}
    \label{fig:Ai-use}
\end{figure}

Figure \ref{fig:Ai-use} provides an overview of the Probe-Monitor-Shape heuristic. 
Table~\ref{tab:mapping} synthesizes the study's three strands by mapping
each Probe-Monitor-Shape element to the strategies that constitute it, the challenges it primarily addresses, and the perceived benefits it helps students realize. The mapping is analytical as it reflects the
correspondence between the descriptions students provided. The mapping is not a per-respondent statistical association, and it is intended to show how
students' regulatory moves function as targeted responses to specific
difficulties. Within this context, the three steps in the probe-monitor-shape heuristic moves form a coherent loop of probing to surface usable output, monitoring to check it, and shaping to constrain it. The heuristic can become a process through which students convert an unpredictable
tool into a source of clearer, more trustworthy, and more time-efficient
support.

It should be noted that many of the observed strategies were improvised and uneven, and this indicates capacity rather than mastery. Students still require structured support to apply these behaviors reliably and effectively. These findings are consistent with existing research that emphasizes the need for learners to develop competencies that enable them to understand and evaluate generative AI systems \citep{kasneci2023chatgpt}.
Furthermore, two reported challenges, \textit{technical limitations} and \textit{domain restrictions}, fall outside this loop. Both challenges are dependent on the LLM-platform, and no reported strategy resolved them. Their absence from the mapping marks the boundary of what individual regulation can accomplish. For example, some difficulties require changes to the tools, institutional guidance, or a student's own study habits rather than better prompting. This distinction reinforces the framing of Probe-Monitor-Shape as a heuristic for managing model output, and not a remedy for every difficulty students encounter.


Prior work often frames LLMs as learning partners \citep{holmes2020artificial, kasneci2023chatgpt}, yet our findings show a different reality where students frequently move through cycles of correction, verification, and steering to make the tool usable. These cycles shift cognitive effort away from the learning task and toward managing the model’s performance. This scenario creates a thin line between learning with the tool and training it when the system fails to meet users’ cognitive expectations. Work on cognitive load and AI‑supported learning challenges shows how this pattern can heighten load, frustration, and mental fatigue in ways that shorten students’ learning span \citep{sweller2005implications}. Hence, we interpret the findings of this study as a descriptive account of self-reported experience rather than a causal claim. The findings also indicate that LLM tools redistribute cognitive and metacognitive effort in ways that may expose disparities in students’ AI literacy, echoing work that positions AI-supported learning as a competency that may disproportionately benefit students who can effectively regulate the technology \citep{Hou2024}.

These observations suggest several directions for practice. Instructors can acknowledge that LLM tools change how students seek and evaluate academic help, and can model how to manage the tool’s unpredictable responses. For example, students can demonstrate verification against course materials, since incorrect and generic output were the difficulties students reported most. Where feasible, courses might offer students structured opportunities to practice probing, monitoring, and shaping alongside technical content, with additional support for students who struggle with verification or contextual reasoning. Because monitoring was the least common move, courses might consider scaffolding support for students to verify and cross-check LLM outputs. Such scaffolding support may offer the greatest leverage for moving students from executive toward instrumental help-seeking. Finally, consistent expectations for appropriate LLM use  across multiple courses, would further help students navigate the tools coherently.


In the future, we will examine how LLM tools can take on more of the cognitive, metacognitive and verification responsibility currently carried by students. This can be explored through self checking features, transparent reasoning steps, and mechanisms that reduce cycles of correction. Additional research is needed to understand how students apply the Probe-Monitor-Shape heuristic across different tasks and contexts, and how LLMs can be designed to better interpret intent, maintain context, and produce stable reasoning. 
Research shows that productive learning is contingent on tools that support thinking rather than simply generating text. Future work will also explore interaction designs that scaffold reasoning, and clarify assumptions and surface alternatives. This will create conditions in which LLMs operate as genuine learning partners.

\bibliography{sn-bibliography}

\end{document}